\documentclass[acmlarge]{acmart-ecaf}

\renewcommand{\baselinestretch}{1.15}

\AtBeginDocument{%
  }

\setcopyright{cc}
\copyrightyear{2026}
\acmYear{2026}
\acmConference[ECAF'26]{Fifth European Conference on Algorithmic Fairness}{September 02--September 04, 2026}{Ghent, BE}

\begin{document}

\title[Semantic Web Framework for FRIA under the EU AI Act]{A Reusable Semantic Web Framework for Evidence-Grounded Fundamental Rights Impact Assessments under the EU AI Act}

\author{Faith Olopade}
\affiliation{%
  \institution{Trinity College Dublin}
  \city{Dublin}
  \country{Ireland}
}
\email{olopadef@tcd.ie}

\author{Delaram Golpayegani}
\affiliation{%
  \institution{ADAPT Centre, Trinity College Dublin}
  \city{Dublin}
  \country{Ireland}
}
\email{golpayes@tcd.ie}

\author{David Lewis}
\affiliation{%
  \institution{ADAPT Centre, Trinity College Dublin}
  \city{Dublin}
  \country{Ireland}
}
\email{dave.lewis@tcd.ie}

\renewcommand{\shortauthors}{Olopade et al.}

\keywords{EU AI Act, Fundamental Rights Impact Assessment, knowledge graph, Semantic Web, LLM classification, algorithmic fairness, public sector AI}

\begin{abstract}
The EU AI Act (Art.~27) requires deployers of high-risk AI systems to conduct
Fundamental Rights Impact Assessments (FRIAs) before deployment, yet the
evidence needed for credible assessments is fragmented across incompatible
incident repositories, risk vocabularies, and legal texts.
We present a reusable Semantic Web-based framework that consolidates this
evidence for two high-risk public sector categories: employment and worker
management (Annex~III(4)) and access to essential public services
(Annex~III(5)(a)).
A curated 150-record corpus is annotated along four axes using keyword, LLM,
and hybrid methods and serialised as a SPARQL-queryable knowledge graph of
1,351 RDF triples.
Five FRIA demonstration scenarios surface 103~records (68.7\% coverage).
Evaluation against a 69-record gold standard reveals that LLM-assisted
classification of the employment domain achieves only $\kappa = 0.045$, a
cautionary result for automated fairness-related evidence retrieval in this
domain.
All artefacts are released openly to support adoption by regulators, national
authorities, and SMEs.
\end{abstract}

\maketitle

\section{Introduction}

Government agencies across Europe are deploying large language models (LLMs)
at scale: welfare eligibility triage, citizen-facing chatbots, recruitment
screening pipelines, and caseworker decision support~\cite{bommasani_foundation_2021}.
The appeal is straightforward, but the risks to fundamental rights are direct.
Bias amplification, hallucinated outputs, and opacity bear on individuals'
access to work and essential services in ways that are not always
recoverable~\cite{weidinger_ethical_2021, bender_dangers_2021}.

Real-world cases demonstrate what inadequate assessment looks like in practice.
In the Netherlands, the Tax Administration's algorithmic risk scoring system
embedded racial profiling and operated without meaningful oversight, wrongly
accusing tens of thousands of families of benefit fraud~\cite{amnesty_childcare_2021}.
The SyRI case established that automated profiling in public welfare contexts
can constitute a violation of the European Convention on Human
Rights~\cite{syri_court_2020, wieringa_syri_accountability_2023}.
In employment, the Mobley v.\ Workday litigation alleges that an AI-powered
r\'{e}sum\'{e} screening tool systematically discriminates on the basis of
age, race, and disability.
A persistent pattern runs through these incidents: high-stakes public sector
AI can cause large-scale rights violations when risk assessment is inadequate.

The EU AI Act responds to this directly.
Article~27 requires that deployers of public sector high-risk AI systems, listed in Annex III, conduct
a Fundamental Rights Impact Assessment (FRIA) before deployment, documenting how the
system may affect rights including non-discrimination, privacy, and good
administration~\cite{eu_ai_act_2024}.
The European Union Agency for Fundamental Rights has noted that many
deployments still lack the tools and evidence base to conduct meaningful
assessments~\cite{fra_assessing_highrisk_2025}.

The evidence deployers need is not available in a usable form: it is fragmented across sources, described inconsistently, and not interoperable, so assembling it for a given assessment demands substantial manual effort. Incident data resides in repositories such as AIAAIC (AI, Algorithmic and Automation Incidents and Controversies) and AIID (AI Incident Database), which use
free-text narratives and community-developed tags rather than AI Act
terminology~\cite{pownall_aiaaic_2024, mcgregor_aiid_2021}.
Existing Semantic Web risk vocabularies---such as Data Privacy Vocabulary (DPV)~\cite{pandit_dpv_v2_2024}, Vocabulary of AI Risks (VAIR)~\cite{vair_vocab_2024}, and AI Risk Ontology (AIRO)~\cite{golpayegani_airo_2022}---provide structured,
machine-readable taxonomies aligned with the regulatory framework but are not
integrated into FRIA workflows~\cite{dpv_2022, vair_vocab_2024}.
Given that legal obligations of the AI Act, including FRIA obligations, and the EU Charter of Fundamental Rights are
expressed in natural language, they do not support automation~\cite{eu_charter_2000}. Therefore,
deployers must \textit{manually} reconcile these disparate sources with no reusable
infrastructure to connect incident evidence, risk vocabularies, and regulatory
obligations.

 Bridging that gap, this paper presents an interoperable framework for AI Act's FRIAs using Semantic Web technologies, with a focus on two categories of Annex III high-risk AI applications.
The full technical implementation is described in the accompanying
dissertation~\cite{olopade_dissertation_2026}.
This extended abstract presents the framework, its key empirical findings,
and their implications for algorithmic fairness research and practice.

\section{Framework}

\subsection{Corpus Construction}

The corpus combines three complementary evidence sources: approximately
100~records from the AIAAIC incident repository~\cite{pownall_aiaaic_2024},
30~entries from the U.S.\ Federal AI Use Case Inventory~\cite{omb_ai_inventory_2024},
and 20~European Court of Human Rights cases from
HUDOC~\cite{hudoc_echr}.
Together these provide 150~records covering LLM-related and analogous AI
deployments in employment and essential public services.
The three sources triangulate across evidence types: AIAAIC captures
real-world failures; the U.S.\ Federal Inventory documents active government
deployments; ECtHR cases provide authoritative legal analysis of rights
violations.

Each record is classified along four axes: (1)~Annex~III high-risk AI domain (employment
or essential services); (2)~implicated EU Charter rights (multi-label);
(3)~risk pattern (bias/discrimination, privacy breach, procedural unfairness,
lack of transparency, or other); and (4)~causal factor (data quality, model
design, deployment context, or oversight failure).

\subsection{Semantic Schema and Knowledge Graph}

The schema is grounded in Semantic Web standards and aligned with
DPV~\cite{dpv_2022, pandit_dpv_v2_2024}, VAIR~\cite{vair_vocab_2024},
AIRO~\cite{golpayegani_airo_2022}, and the FRIA
ontology~\cite{rintamaki_pandit_fria_ontology_2025}.
Reusing established vocabularies reduces the learning burden for adopters
and maximises interoperability with existing compliance tooling.
The annotated corpus is serialised in two formats: a Turtle RDF knowledge
graph of 1,351~triples (193~nodes, 965~edges) enabling information retrieval using the SPARQL query language\footnote{\url{https://www.w3.org/TR/sparql11-query/}},
and 150~JSON-LD records for web-compatible consumption.\footnote{All schema artefacts, the knowledge graph, and the SPARQL queries are available at \url{https://github.com/faitholopade/Dissertation}.}

\subsection{Annotation Pipeline}

Three annotation methods are implemented and compared.
A keyword baseline uses synonym-expanded dictionary matching against
controlled vocabularies.
An LLM classifier applies Claude Sonnet~\cite{anthropic_claude3_2024} with
zero-temperature sampling, few-shot prompting, and structured output
constraints.
A hybrid method resolves conflicts between keyword and LLM outputs using
priority logic, retaining keyword classifications for domain and LLM
classifications for rights where each method performs better.

\subsection{Regulatory Crosswalk}

A regulatory crosswalk maps the obligations arising from Annex~III(4)
and Annex~III(5)(a) to specific articles of the EU Charter of Fundamental
Rights~\cite{eu_charter_2000}.
This mapping is not enumerated in the AI Act itself; the crosswalk makes it
explicit and machine-readable, reducing interpretive uncertainty for deployers
and providing the structured link between regulatory requirements and rights
evidence that the schema's multi-label rights axis builds on.

\subsection{Query Interface}

A Flask web application loads the knowledge graph on startup and allows users
to filter records by Annex~III domain, Charter rights, and risk pattern
through a form-based interface, moving the framework toward the interactive
compliance tooling anticipated by Art.~27(5)~\cite{eu_ai_act_2024}.

\section{Evaluation}

\subsection{Gold Standard and Agreement Metrics}

Of the 150~corpus records, 69 (46\%) were manually annotated to form a gold
standard, stratified across sources and both Annex~III domains.
Each record was annotated by reviewing original source material rather than
summary fields alone, following written guidelines with explicit decision
criteria per axis.
Cohen's $\kappa$~\cite{cohen_kappa_1960} and percentage agreement are
reported for each axis and method, interpreted using the Landis and Koch
scale~\cite{landis_koch_1977}.

The 69~manually annotated records (46\%) are a deliberately stratified gold
standard rather than a partial annotation effort. The framework's purpose is to
test whether automated annotation can scale beyond what manual labelling
feasibly covers, so the manual subset is sized to give reliable per-axis
agreement estimates across both Annex~III domains and all three sources, not to
annotate the corpus exhaustively. Manually labelling all 150~records would
defeat the evaluation, since the pipeline exists precisely to avoid that cost at
scale.

\subsection{Domain Classification}

For the essential services domain, the best-performing method achieves
$\kappa = 0.525$ (moderate agreement).
For employment, the best-performing method achieves $\kappa = 0.045$---near-chance.
This divergence is attributable to spurious vocabulary correlations: a
disproportionate share of records discuss employment incidentally rather than
as the primary domain of harm, and both keyword and LLM classifiers
over-predict employment for records containing general labour market language.

The implication for algorithmic fairness research is direct.
Employment discrimination is among the most extensively documented harms of
automated decision-making, and Annex~III(4) covers the AI Act's highest-risk
employment applications.
Near-chance agreement means that automated annotation of employment-domain
risk evidence cannot currently be trusted without human review.
Deployers relying on such tools to surface employment risk evidence for a FRIA
risk generating an evidence base that is both incomplete and misleading.
The finding holds across models: a comparison with GPT-4o-mini yields
$\kappa = 0.196$ for domain classification, confirming the difficulty lies
in the task rather than in a single model.

\subsection{Coverage and Risk Patterns}

\begin{figure}[t]
  \centering
  \includegraphics[width=\linewidth]{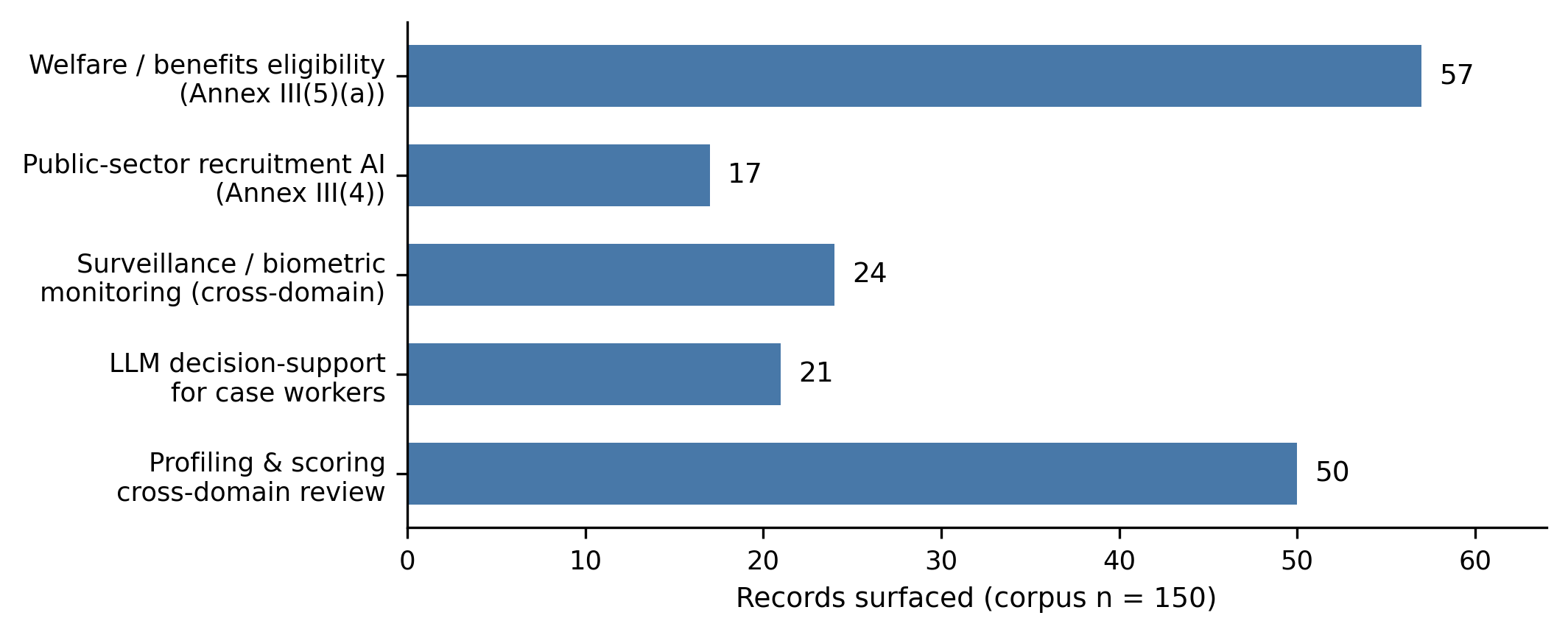}
  \caption{Records surfaced by each of the five FRIA demonstration scenarios.
  Scenarios may retrieve overlapping records; their union is 103 of the 150
  corpus records (68.7\% coverage). The employment recruitment scenario surfaces
  the fewest records, mirroring the near-chance employment domain
  classification.}
  \Description{Horizontal bar chart of the number of corpus records surfaced by
  each of the five FRIA demonstration scenarios.}
  \label{fig:coverage}
\end{figure}

Five FRIA demonstration scenarios query the populated knowledge graph for
records relevant to realistic deployer contexts: welfare eligibility AI,
public sector recruitment screening, surveillance in public housing, LLM
decision-support for caseworkers, and a cross-domain thematic review by
a national regulator.
Across all five, 103 of the 150~records are surfaced (68.7\% coverage).
The 31.3\% not surfaced are predominantly records with unknown classifications
produced by pipeline underclassification---the direct consequence of the
domain agreement failures described above. Figure~\ref{fig:coverage} reports
per-scenario retrieval: the welfare-eligibility and cross-domain profiling
scenarios surface the largest evidence pools (57 and 50 records), while the
employment-focused recruitment scenario surfaces the fewest (17), consistent
with the near-chance employment agreement reported above.

The hybrid method reduces unknown risk pattern classifications from 61.3\%
(keyword baseline) to 14.0\%, substantially enlarging the evidence base
available for compliance queries.
Agreement on risk patterns remains in the slight-to-fair range, with
procedural unfairness and lack of transparency the categories of greatest
disagreement, driven primarily by multi-label harm structures where incidents
involve co-occurring failure modes.

\subsection{Use Case: A Welfare Eligibility Deployer}

A public body preparing to deploy an AI system that supports social-welfare
eligibility decisions (Annex~III(5)(a)) must document, before deployment, how
the system may bear on the rights to social security (Art.~34) and
non-discrimination (Art.~21). Using the query interface, the deployer filters
the knowledge graph for essential-services records implicating these rights; the
query returns 57 candidate records (Scenario~A in Figure~\ref{fig:coverage}).
Each record carries its risk-pattern and causal-factor classifications together
with full provenance back to the originating incident, so a flagged
non-discrimination risk can be traced to specific precedents---such as the Dutch
childcare-benefits scandal~\cite{amnesty_childcare_2021} and the SyRI profiling
case~\cite{syri_court_2020,wieringa_syri_accountability_2023}---rather than to a
general assertion that such risks exist. This is what separates an
evidence-grounded FRIA from a narrative one: the deployer can judge whether
documented failures are analogous to its own context, and the machine-readable
output lets a supervisory authority later verify that comparable deployers
considered the same rights, as anticipated by Art.~27(5).

\section{Limitations}

Four limitations bear on the framework's current scope.
The gold standard was produced by a single annotator; multi-annotator
reliability using Fleiss'~$\kappa$~\cite{fleiss_kappa_1971} is a necessary
next step before pipeline outputs can be used without human review.
The corpus covers only two of the eight Annex~III categories; the schema is
designed for extensibility but coverage of biometrics, law enforcement, and
migration is not yet tested, a gap that prior work on high-risk AI
classification highlights~\cite{golpayegani_highrisk_2024,
golpayegani_highrisksemacl_facct_2023}.
AIAAIC is predominantly English-language and draws heavily on U.S.\ and
U.K.\ sources, introducing geographic bias in an EU regulatory context.
Finally, employment classification quality is insufficient for unsupervised
deployment and requires human oversight for any FRIA relying on that
domain's evidence.

\section{Relevance to ECAF}

This work contributes across three of ECAF's disciplinary areas.

\textbf{Policy and Law.}
The regulatory crosswalk and FRIA demonstration scenarios directly address
impact assessment obligations under the AI Act, providing concrete,
reusable infrastructure for Art.~27 compliance and contributing to emerging
standardisation of AI incident
reporting~\cite{oecd_ai_incidents_framework_2025}. This work can also be considered as a foundation for the automated tool the AI Office should develop for FRIA, as per Art. 27(5). 
The analysis of GDPR versus AI Act impact assessment
requirements~\cite{rintamaki_impact_2026} provides complementary legal
framing.

\textbf{Computer Science.}
The annotation pipeline, multi-method comparison, and gold standard evaluation
contribute empirical evidence on the reliability of LLM-assisted
classification in a regulatory context, with direct implications for auditing
framework design.

\textbf{Social Sciences.}
The corpus documents real-world AI failures in employment and essential
services, two domains where automated systems bear most directly on the
rights of marginalised groups.
The near-chance employment domain agreement ($\kappa = 0.045$) is a
cautionary result for practitioners building automated fairness assessment
tools in this domain.

\section{Conclusion}

We have presented a Semantic Web-based framework that consolidates fragmented
AI risk evidence into a structured, SPARQL-queryable knowledge graph supporting
EU AI Act FRIA compliance.
Five FRIA demonstration scenarios achieve 68.7\% coverage of a 150-record
corpus, and gold standard evaluation reveals that LLM-assisted employment
domain classification is near-chance---a result with direct implications for
automated fairness assessment in the domain most associated with algorithmic
discrimination.
All schema artefacts and annotated data are released under CC~BY~4.0 and
pipeline source code under the MIT Licence at
\url{https://github.com/faitholopade/Dissertation}, to support adoption by
regulators, national authorities, and SMEs extending the framework to further
Annex~III categories.

\begin{acks}
This work was supported by the School of Computer Science and Statistics,
Trinity College Dublin.
The authors acknowledge the support of the ADAPT Centre for Digital Content
Technology, funded under the Science Foundation Ireland Research Centres
Programme (\#13/RC/2106\_P2).
\end{acks}

\bibliographystyle{ACM-Reference-Format}
\bibliography{references}

\end{document}